\documentclass[11pt,twoside]{article}
\usepackage{asp2010}

\resetcounters

\begin{document}

\title{On the Distance and Age of the Pulsar Wind Nebula 3C\,58}
\author{Roland Kothes$^1$
\affil{$^1$National Research Council Canada, Herzberg Institute of Astrophysics,
Dominion Radio Astrophysical Observatory, P.O. Box 248, Penticton, BC V2A~6J9, Canada}}

\begin{abstract}
There is a growing community of astronomers
presenting evidence that the pulsar wind nebula 3C\,58 is much older than the 
connection with the historical supernova of A.D.~1181 would indicate. Most of the 
strong evidence against a young age for 3C\,58 relies heavily on the assumed 
distance of 3.2~kpc determined with \ion{H}{i} absorption measurements. I have revisited
this distance determination based on new \ion{H}{i} data from the Canadian Galactic 
Plane Survey and added newly determined distances to objects in the neighbourhood, 
which are based on direct measurements by trigonometric parallax. This leads to 
a new more reliable distance estimate of 2~kpc for 3C\,58 and makes the 
connection between the pulsar wind nebula and the historical event from 
A.D.~1181 once again much more compelling.
\end{abstract}

\section{Introduction}

Of all the supernova remnants (SNRs) which are linked to historically observed 
supernova events the connection between the pulsar wind nebula (PWN) 3C\,58 and 
the supernova explosion observed A.D.~1181 by Chinese and Japanese astronomers 
is probably the most disputed one. \citet{step71} claimed that there is a high 
probability for a connection between the Guest Star from A.D.~1181 and the 
supernova explosion that created 3C\,58. This has been revisited and confirmed 
by \citet{step02}. The main arguments are the length of the visibility of the
1181 event, indicating a supernova explosion, and the lack of any other supernova remnant 
candidates. For such a recent supernova there should
be a visible SNR. Although it is very difficult to argue with this argument,
strong evidence against such a young age for 3C\,58 has been mounting 
\citep[for a list of the main arguments see e.g.][Table 3]{fese08}. 

Most arguments against a young age for 3C\,58 rely heavily on the 
assumed distance of 3.2~kpc \citep{robe93}. This distance was determined 
kinematically from \ion{H}{i} absorption measurements by comparing the 
resulting radial velocity with a flat rotation curve for the Outer Galaxy. 
However, in particular for Perseus arm objects, this
leads to a significant overestimate of the distance. A 
spiral shock in the Perseus arm is ``pushing'' objects towards us, 
giving them a higher negative radial velocity, which makes them appear to 
be farther away \citep{robe72}. Examples for this effect on 
distances for SNRs and PWNe can be found 
in \citet{koth02b,koth03} and \citet{fost04}. SNRs in particular rely on
kinematic distance estimates, since their distances cannot be easily determined
directly or by related stars as in the case of \ion{H}{ii} regions.  

In this short article I present a new distance estimate for 3C\,58 and
based on this new distance I will discuss and re-evaluate evidence presented in 
the literature against the historical connection of 3C\,58. I will show that the 
new distance changes the
characteristics of this PWN quite dramatically, which leads to a higher 
probability for its historical connection.

\section{Observations}

The \ion{H}{i} data I used for the determination of the new absorption profile towards
3C\,58 were obtained with the Synthesis Telescope at the Dominion Radio
Astrophysical Observatory \citep[DRAO,][]{DRAO} as part of the Canadian Galactic Plane
Survey \citep[CGPS,][]{tayl03}. Single antenna data are incorporated into the 
interferometer maps to ensure accurate representation from the largest
structures down to the resolution limit of about 1\arcmin. The low spatial
frequency data was drawn from the Low-Resolution DRAO Survey of the CGPS region,
which was observed with the 26-m radio telescope at DRAO \citep{higg00}. The
resolution in the final maps varies slightly across the final images close to
$1\arcmin \times 1\arcmin {\rm cosec(DEC)}$. At the centre of 3C\,58 the 
resolution of the \ion{H}{i} data is $58\arcsec \times 65\arcsec$ at an angle of 
$75\deg$ for the major axis (counter-clockwise from the Galactic longitude 
axis). The RMS noise 
is about 3~K $T_B$ in each velocity channel of width 0.82446~km~s$^{-1}$ at
a velocity resolution of 
1.3~km~s$^{-1}$.

\section{Results}

\subsection{\ion{H}{i} Absorption and Emission Associated with 3C\,58}

The latest distance
determination for 3C\,58 by \ion{H}{i} absorption measurements resulted in a 
systemic velocity of
$\sim -38$~km~s$^{-1}$ and a Perseus arm location \citep{robe93}. This was translated 
with a flat rotation model for our Galaxy
and the IAU supported values for the Sun's Galacto-centric distance of 
$R_\odot = 8.5$~kpc and the Sun's circular motion around the Galactic centre of
$v_\odot = 220$~km\,s$^{-1}$ to a distance of 3.2~kpc.

\begin{figure}[!ht]
\centerline{
  \includegraphics[bb = 100 263 475 620,width=0.75\textwidth,clip]
  {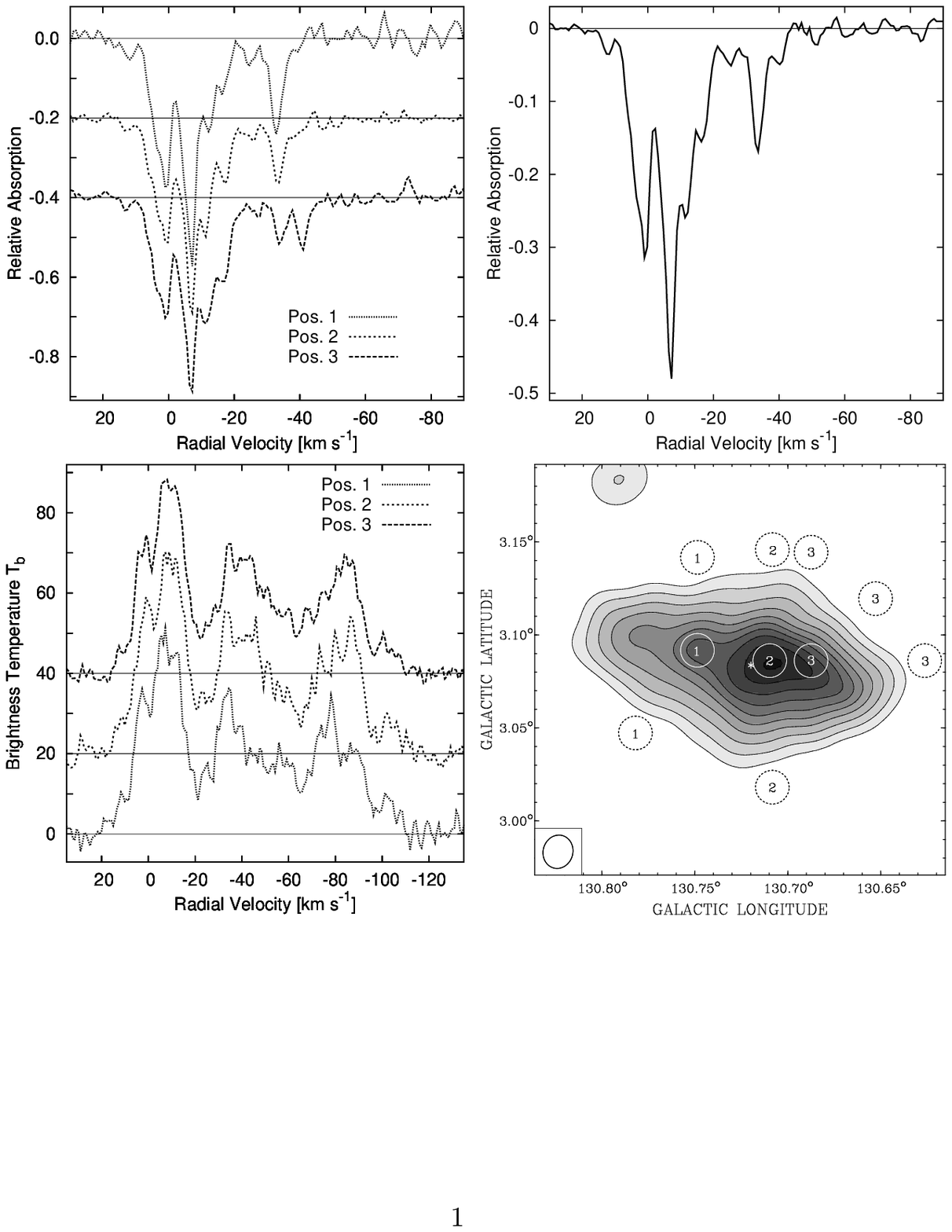}}
 \caption{{\bf Top}: \ion{H}{i} absorption profiles of 3C\,58. The relative absorption 
     of the radio continuum signal at 1420~MHz is displayed as a function of radial 
     velocity $v_{LSR}$. Left: Absorption profiles of 3 different positions are 
     displayed as indicated in the bottom right panel. Right: Profile
     calculated using all pixels on 3C\,58 with $T_b \ge 100$\,K. Each pixel was 
     weighted by its intensity.
     {\bf Bottom}: Left: \ion{H}{i} emission profiles used to determine the 
     absorption profiles in the top left. These were averaged
     over the positions marked in the right panel. Right: Continuum image of
     3C\,58 taken from the CGPS. The "on" and "off" positions used to calculate the 
     \ion{H}{i} absorption profiles are indicated by solid and dashed circles,
     respectively.}
   \label{abs}
\end{figure}

\begin{figure}[!ht]
\centerline{
 \includegraphics[bb = 35 129 553 475,width=0.85\textwidth,clip]
 {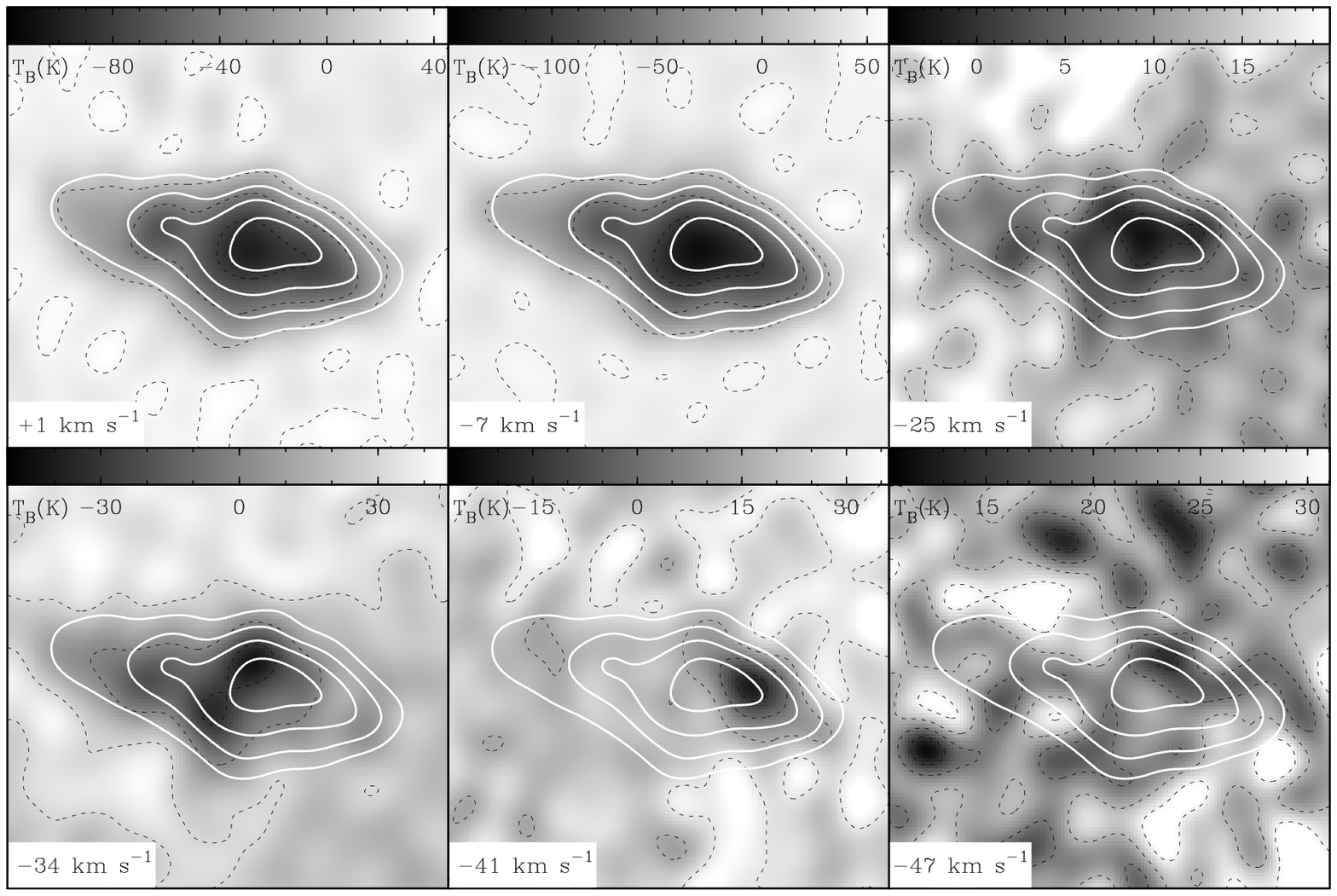}}
\caption{\ion{H}{1} channel maps of 3C\,58 taken at the peaks of the 
absorption spectra at +1 (top left) -7 (top centre), -34 (bottom left), and
-41~km\,s$^{-1}$ (bottom centre). The right panels show 10 channels averaged over
the interarm velocities between the Local and Perseus arm (top) and 10 channels
averaged over the velocities just outside the last absorption feature.
The continuum emission of 3C\,58 is indicated
by the white contours. The black dashed lines represent the \ion{H}{i} contours at the 
levels indicated by the labels of the colour bars.}
\label{abs2}
\end{figure}

I used the CGPS data to determine a new \ion{H}{i} absorption profile towards 3C\,58
to study not only the integrated absorption spectrum, but also the changes over 
the PWN. The resulting profiles are shown in Fig.~\ref{abs}. \ion{H}{i} channel maps
taken at the peaks of the absorption spectrum and channels averaged over the 
interarm areas
are displayed in Fig.~\ref{abs2}. The integrated
absorption spectrum is literally identical to the one published by 
\citet{robe93}. However, I found that the weak peak, seen
at around $-41$~km~s$^{-1}$ is actually a real absorption feature. It is not 
seen over the entire PWN, but only on the right hand side at position 3 
(see Figs.~\ref{abs} and \ref{abs2}).
This is likely the reason why this feature was not obvious in previously
published integrated absorption spectra.

\begin{figure}[!ht]
\centerline{
 \includegraphics[width=0.85\textwidth]
 {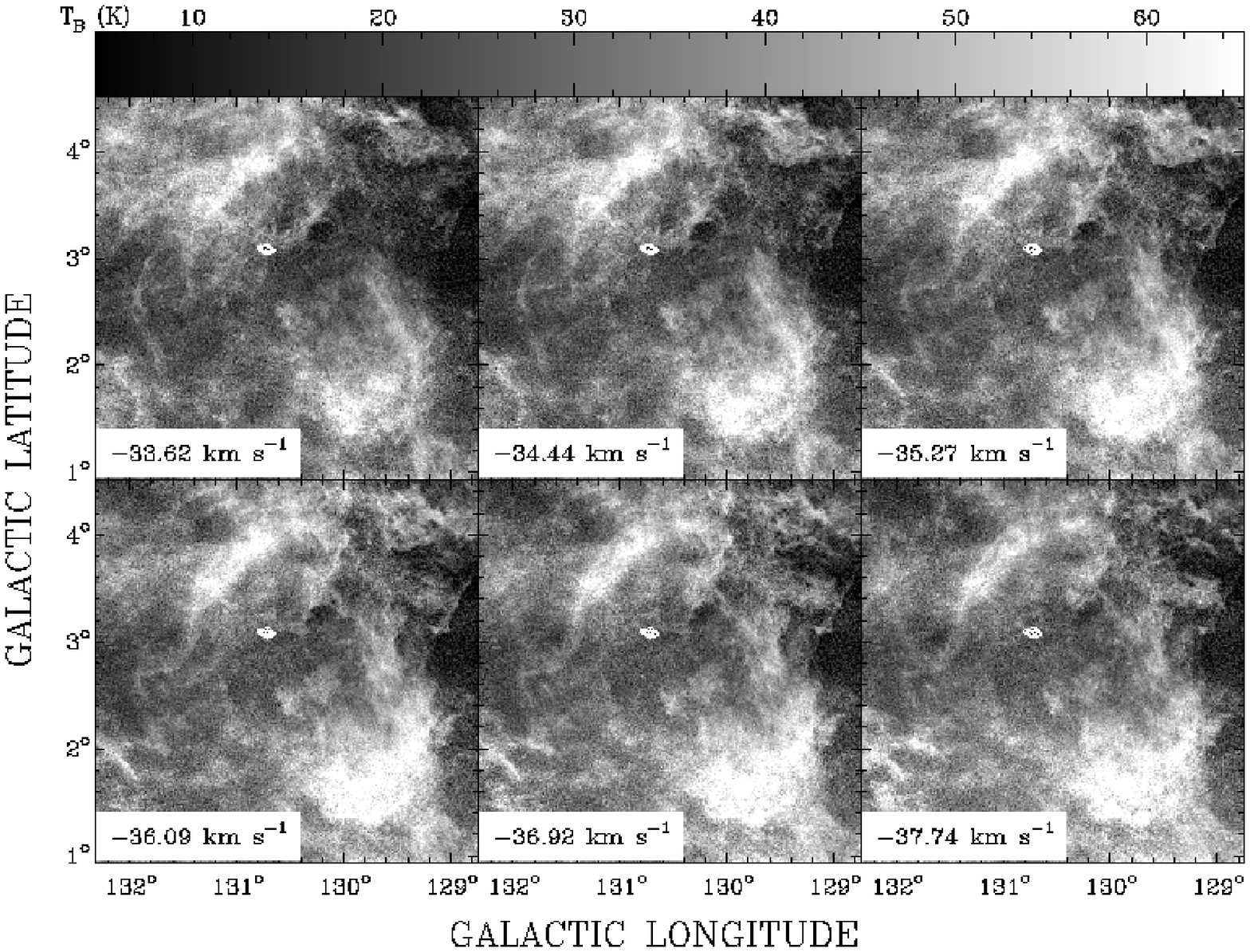}}
\caption{\ion{H}{1} channel maps of the area around 3C\,58. Each channel is 
0.82446~km\,s$^{-1}$ wide and the centre velocity is indicated. The location 
and emission structure of 3C\,58 are marked by the white contours.}
\label{hiemi}
\end{figure}

\citet{wall94} discussed the possible association of 3C\,58 with a large
interstellar bubble. An inspection of the CGPS
\ion{H}{i} data set reveals that this double shell structure is actually a very smooth
object (see Fig.~\ref{hiemi}). The bubble is visible from about 
$-33$ to $-38$\,km\,s$^{-1}$ and 3C\,58 is located in the top left area of it.
If 3C\,58 was located within
this bubble, the last absorption feature at about $-41$\,km\,s$^{-1}$ is likely
produced by the shell of the bubble expanding towards us. 

\subsection{The Systemic Velocity of 3C\,58}

\citet{robe93} found a Perseus arm location at a systemic velocity of about 
$v_{LSR} = -38$\,km\,s$^{-1}$, which they translated to a distance of 3.2\,kpc.
This systemic velocity was confirmed with the discovery of the large bubble 
by \citet{wall94}. The CGPS \ion{H}{i} data (Figs.~\ref{abs} and \ref{abs2}) show an 
additional absorption feature at about $-41$\,km\,s$^{-1}$ and a systemic 
velocity of about $-36$\,km\,s$^{-1}$ for the bubble. There is no absorption
at velocities beyond $-41$\,km\,s$^{-1}$. In Fig.\ref{abs2} the map averaged
around $-25$\,km\,s$^{-1}$, representative of the interarm between the Local arm and
the Perseus arm shows significant absorption, however, the map averaged over velocities 
beyond $-41$\,km\,s$^{-1}$ is free of absorption even though this velocity range
shows brighter emission than the interarm around $-25$\,km\,s$^{-1}$. This confirms the
Perseus arm location for 3C\,58. Presumably 3C\,58 is
located inside the bubble shown in Fig.~\ref{hiemi} and the last absorption feature,
which is only partly visible in the absorption of 3C\,58 is caused by a part of that bubble moving
towards us. That would give this bubble
an expansion velocity of about 5\,km\,s$^{-1}$. Using the equations from
\citet{mcgriff02a} we calculate a dynamic age of about $3\times 10^6$\,yr
assuming it is the result of a single supernova explosion and $6\times 10^6$\,yr for
a stellar wind bubble.

\section{Discussion}

\subsection{The Distance to 3C\,58}

\begin{figure}[!ht]
\centerline{
 \includegraphics[bb = 22 18 616 585,width=0.50\textwidth,clip]
 {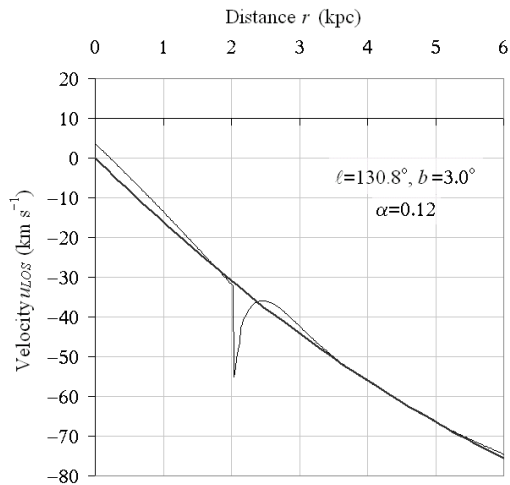}}
\caption{Plot of the distance-velocity relation in the direction of 3C\,58
determined with the method by\citet{fost06}.}
\label{dvplot}
\end{figure}

The method by \citet{fost06} was used to determine a distance-velocity relation in
the direction of 3C\,58 (Fig.~\ref{dvplot}). For a
systemic velocity of $-36$\,km\,s$^{-1}$ there are two possible distances: in the
Perseus arm shock at about 2\,kpc and beyond Perseus arm at about 2.5 - 2.8\,kpc.
Interstellar material is compressed in the spiral shock, forms molecules, and then
stars. It takes a long times -- several $10^7$ years -- to migrate beyond the Perseus arm.
Therefore, considering the dynamical age of the bubble and the high probability that
the progenitor star of 3C\,58 formed in the spiral shock, the likelihood of 3C\,58 being 
at the farther distance is very low.

There is a second independent method to determine a distance to 3C\,58, to relate
the PWN to a nearby object with a more reliable distance estimate. 
The W\,3/4/5 \ion{H}{ii} region
complex and the related SNR HB\,3 are just a few degrees away from 3C\,58. Both, HB\,3
and W\,3/4/5, have very similar systemic velocities \citep[e.g.][]{rout91}. They are all 
located in the Perseus arm and for W\,3 the distance was determined with trigonometric 
parallax to related masers to be $1.95\pm0.04$\,kpc and $2.04\pm0.07$\,kpc 
by \citet{xu06} and \citet{hach06}, respectively. This nicely agrees with the 2\,kpc
I estimated for 3C\,58 from \ion{H}{i} absorption.

\subsection{The Age of 3C\,58}

There are a number of arguments against a relation between the PWN 3C\,58 and
the historical supernova explosion from A.D. 1181 as outlined in
\citet[Table 3]{fese08}. I will discuss the two major groups of these arguments below.

\subsubsection{Expansion Studies}

One important aspect of expansion studies that is not always fully appreciated is the
fact that a supernova explosion is a one-time event. For any expansion
study, observed features related to the explosion could have been decelerated but not 
accelerated. On the other hand, this
may not be true for synchrotron emitting filaments related to a PWN, because those have 
a continuous source of energy in the central pulsar. Hence, the interpretation of 
expansion studies of synchrotron filaments is not straightforward.

A comparison of the two major expansion studies in optical 
\citep[$\Rightarrow$ age $t \approx 3500$\,yr]{fese08} and radio 
\citep[$\Rightarrow t \approx 7000$\,yr]{biet06} 
already implies significant deceleration for the radio structures relative to the optical 
filaments unless the energy released by the pulsar significantly accelerated all the 
supernova ejecta. A comparison of the typical $10^{51}$\,erg released in a supernova
and the approximate energy released by the pulsar since birth of $\sim 10^{48}$\,erg
\citep{chev04} negates that possibility.

The optical filaments are created by material accelerated through the supernova 
explosion. Because those filaments could have been decelerated but not accelerated 
a simple averaging of the expansion velocity of individual filaments would not 
necessarily lead to a good age estimate unless the scatter is entirely produced 
by uncertainties in the observations or systematic errors, which is certainly not the case in the study
by \citet{fese08}. The fastest filament should be taken, since it presumably shows the 
lowest deceleration. There is quite 
a wide spread in velocity in the optical study by \citet{fese08}, which again
indicates that a lot of the emitting material must have been 
decelerated. Therefore the assumption of insignificant deceleration for the
optical filaments is invalid as well.
This significantly  weakens the case for a large age based on the optical and 
radio expansion studies. 

\subsubsection{PWN Evolution and Energy}

\citet{chev04} discussed theoretical calculations of the evolutionary path of a PWN and 
compared their results for swept up mass and internal energy with those from 
observations. Both of these results rely heavily on the assumed distance. 
\citet{bocc01} determined from their X-ray observations a swept-up mass
of $M_{sw} = 0.1 d_{3.2}^{2.5}$\,M$_\odot$, assuming a radius $R$ of $2.5\arcmin$, which results in 0.1\,M$_\odot$ for a
distance $d$ of 3.2\,kpc. The theoretical value from \citet{chev04} equates to:
$M_{sw} = \dot{E} R^{-2} t^3$ ($\dot{E}$: rotational energy loss rate of the pulsar) 
resulting in 0.005\,M$_\odot$. Even with the relatively high
uncertainty of this kind of calculation the large discrepancy suggests a much larger 
age for 3C\,58. To increase the theoretical value to the observed one the age of the
PWN has to be about $t \approx 2300$\,yr. The other argument is that the minimum energy
required to produce the synchrotron nebula is about $10^{48}$\,erg from equipartition
considerations. This value depends on the radio luminosity and thereby on
$d^2$. The total energy released from the pulsar into the nebula, however, which
can be approximated by $\dot{E} t$ is about $0.7\times 10^{48}$\,erg, which is much smaller.
This value is distance independent.

For a distance of 2\,kpc the minimum energy required to produce the synchrotron emission
decreases to $0.4\times 10^{48}$\,erg, which is now lower than the total energy released by
the pulsar, which remains unchanged. The values for the swept up masses equate to 0.030\,M$_\odot$ and
0.013\,M$_\odot$ for the X-ray observations and the theoretical calculations, respectively.
The two results for the swept-up mass still differ but not by much. To get those values to
be equal requires either an age of about 1100\,yr or a distance of 1.7\,kpc. For
a distance of 1.9\,kpc the required age would be reduced to 1000\,yr.

\section{Conclusion}

I derived a new more reliable distance of 2\,kpc to the PWN 3C\,58, by means of 
\ion{H}{i} absorption in combination with a newly determined distance-velocity relation 
and by relating this PWN to a nearby \ion{H}{ii} region SNR complex of well known 
distance. This new distance changes many characteristics of this PWN quite
dramatically which once again makes the connection between 3C\,58 and the
historical supernova explosion from 1181 A.D. much more likely.

\acknowledgements

I would like to thank Tyler Foster for providing me with a distance-velocity
diagram in the direction of 3C\,58 using his modeling of Galactic hydrogen 
distribution. The Dominion Radio Astrophysical Observatory is a National Facility
operated by the National Research Council. The Canadian Galactic
Plane Survey is a Canadian project with international partners, and is
supported by the Natural Sciences and Engineering Research Council
(NSERC).

\bibliography{kothes_roland}

\end{document}